\documentclass[fleqn,12pt,twoside]{article}
\usepackage{espcrc1}

\usepackage{epsfig}
\usepackage[figuresright]{rotating}

\newcommand{\AmS}{{\protect\the\textfont2
  A\kern-.1667em\lower.5ex\hbox{M}\kern-.125emS}}

\title{Identified particles in Au+Au collisions at $\sqrt{s_{_{NN}}}$ = 200 GeV}

\author{Barbara Wosiek for the PHOBOS Collaboration }%

\begin{document}

% typeset front matter
\maketitle
\noindent
{\small
B.B.Back$^1$,
M.D.Baker$^2$,
D.S.Barton$^2$,
R.R.Betts$^6$,
M.Ballintijn$^4$,
A.A.Bickley$^7$,
R.Bindel$^7$,\\
A.Budzanowski$^3$,
W.Busza$^4$,
A.Carroll$^2$,
M.P.Decowski$^4$,
E.Garc\'{\i}a$^6$,
N.George$^{1,2}$,\\
K.Gulbrandsen$^4$,
S.Gushue$^2$,
C.Halliwell$^6$,
J.Hamblen$^8$,
G.A.Heintzelman$^2$,
C.Henderson$^4$,\\
D.J.Hofman$^6$,
R.S.Hollis$^6$,
R.Ho\l y\'{n}ski$^3$,
B.Holzman$^2$,
A.Iordanova$^6$,
E.Johnson$^8$,
J.L.Kane$^4$,
J.Katzy$^{4,6}$,
N.Khan$^8$,
W.Kucewicz$^6$,
P.Kulinich$^4$,
C.M.Kuo$^5$,
W.T.Lin$^5$,
S.Manly$^8$,
D.McLeod$^6$,
J.Micha\l owski$^3$,
A.C.Mignerey$^7$,
R.Nouicer$^6$,
A.Olszewski$^3$,
R.Pak$^2$,
I.C.Park$^8$,
H.Pernegger$^4$,
C.Reed$^4$,
L.P.Remsberg$^2$,
M.Reuter$^6$,
C.Roland$^4$,
G.Roland$^4$,
L.Rosenberg$^4$,
J.Sagerer$^6$,
P.Sarin$^4$,
P.Sawicki$^3$,
W.Skulski$^8$,
S.G.Steadman$^4$,
P.Steinberg$^2$,
G.S.F.Stephans$^4$,
M.Stodulski$^3$,\\
A.Sukhanov$^2$,
J.-L.Tang$^5$,
R.Teng$^8$,
A.Trzupek$^3$,
C.Vale$^4$,
G.J.van~Nieuwenhuizen$^4$,
R.Verdier$^4$,
B.Wadsworth$^4$,
F.L.H.Wolfs$^8$,
B.Wosiek$^3$,
K.Wo\'{z}niak$^3$,
A.H.Wuosmaa$^1$,
B.Wys\l ouch$^4$\\}

%\vspace{2mm}
\noindent
{\small 
$^1$~Argonne National Laboratory,
$^2$~Brookhaven National Laboratory,
$^3$~Institute of Nuclear Physics, Krak\'{o}w, Poland,
$^4$~Massachusetts Institute of Technology,
$^5$~National Central University, Chung-Li, Taiwan,
$^6$~University of Illinois at Chicago,
$^7$~University of Maryland,
$^8$~University of Rochester}

\begin{abstract}
The yields of identified particles have been measured at RHIC for Au+Au collisions
at $\sqrt{s_{_{NN}}}$ = 200 GeV using the PHOBOS spectrometer. The ratios of
antiparticle to particle yields near mid-rapidity are presented. 
The first measurements of the invariant yields of charged pions, kaons and protons
at very low transverse momenta are also shown.
\end{abstract}

\section{INTRODUCTION}

In the PHOBOS experiment, particles are measured and identified in a two-arm
spectrometer \cite{spec} consisting of layers of Si detectors placed 
on either side of the beam axis in a 2 Tesla magnetic field.  
Unique features of the PHOBOS spectrometer, 
such as a close proximity
of the sensitive detector layers to the interaction region, little
material between the collision vertex and Si layers and high segmentation of Si
detectors, permit
precise measurements of particles down to very small transverse momenta.

In this paper we present the measured ratios of charged antiparticles
to particles for pions, kaons and protons produced near mid-rapidity. These
measurements provide information on  properties of the system at the
chemical freeze-out point and on baryon transport processes. The 
invariant yields of charged particles at very low transverse
momenta are also shown. 
The results of these measurements are sensitive to
large-volume
physics phenomena and to the effects of collective radial expansion of the system.
An enhancement of the production of very low transverse momentum particles could indicate
new long distance scale physics in heavy ion collisions \cite{busza}.

\section{ANTIPARTICLE TO PARTICLE RATIOS}

The ratio of negatively to positively charged particles was measured 
during the RHIC 2001 Run
using spectrometer data taken with both polarities
of the magnetic field.
Event triggering and determination of the collision centrality was
provided by two sets of scintillator paddle counters \cite{trigcent}.
 The rapidity
coverage for the ratio measurements extends from about 0.2 to 0.8
for kaons and protons and 0.35 to 1.3 for pions. More details on the tracking,
particle identification and details of the ratio measurements can be 
found elsewhere \cite{ratio1,ratio2}.

The fully corrected ratios measured within our rapidity acceptance 
for the 12\% most central Au+Au collisions
at 200 GeV  are \cite{ratio2}: 
$<\pi^->/<\pi^+> = 1.025 \pm 0.006(stat.) \pm
0.018(syst.)$, $<K^->/<K^+> = 0.95 \pm 0.03(stat.) \pm 0.03(syst.)$ and
$<\overline{p}>/<p> = 0.73 \pm 0.02(stat.) \pm 0.03(syst.)$. The $<\overline{p}>/<p>$ 
ratio increases by about 25\% over the value at $\sqrt{s_{_{NN}}}$ = 130 GeV \cite{ratio1},
indicating a rapidly decreasing net-baryon
density near mid-rapidity.
The estimated baryochemical potential ($\mu_{B}$) for the system formed in 
central Au+Au collisions at 200 GeV is
$\mu_{B}=27 \pm 2$ MeV \cite{ratio2}.

\begin{figure}[htb]
\begin{minipage}[t]{80mm}
%\framebox[79mm]{\rule[-26mm]{0mm}{52mm}}
\epsfig{file=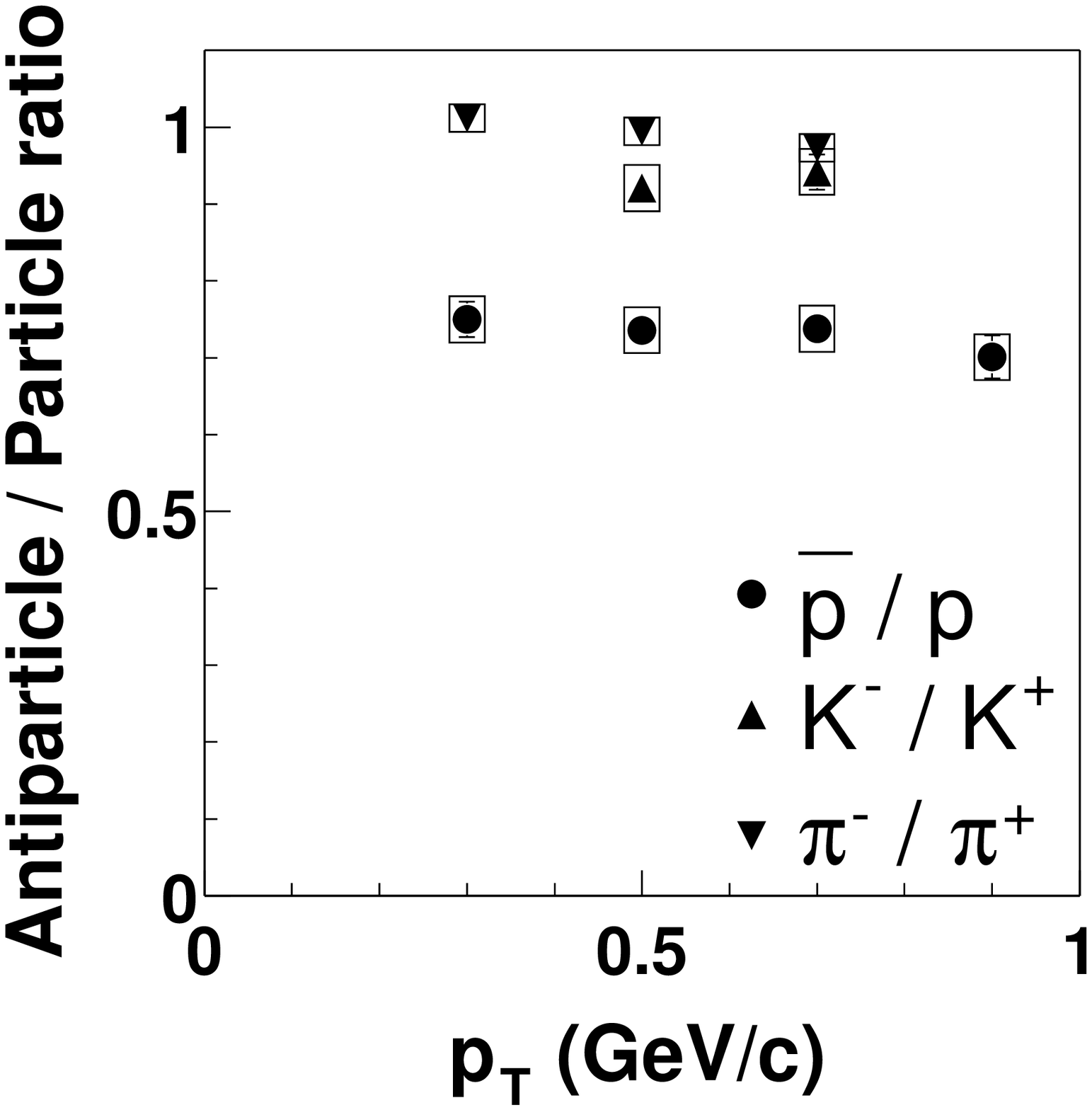,width=65mm}
\caption{Antiparticle to particle ratios as a function of transverse
momentum for selected central collisions. Boxes denote systematic errors.}
\label{fig:largenenough}
\end{minipage}
\hspace{\fill}
\begin{minipage}[t]{75mm}
%\framebox[74mm]{\rule[-26mm]{0mm}{52mm}}
\epsfig{file=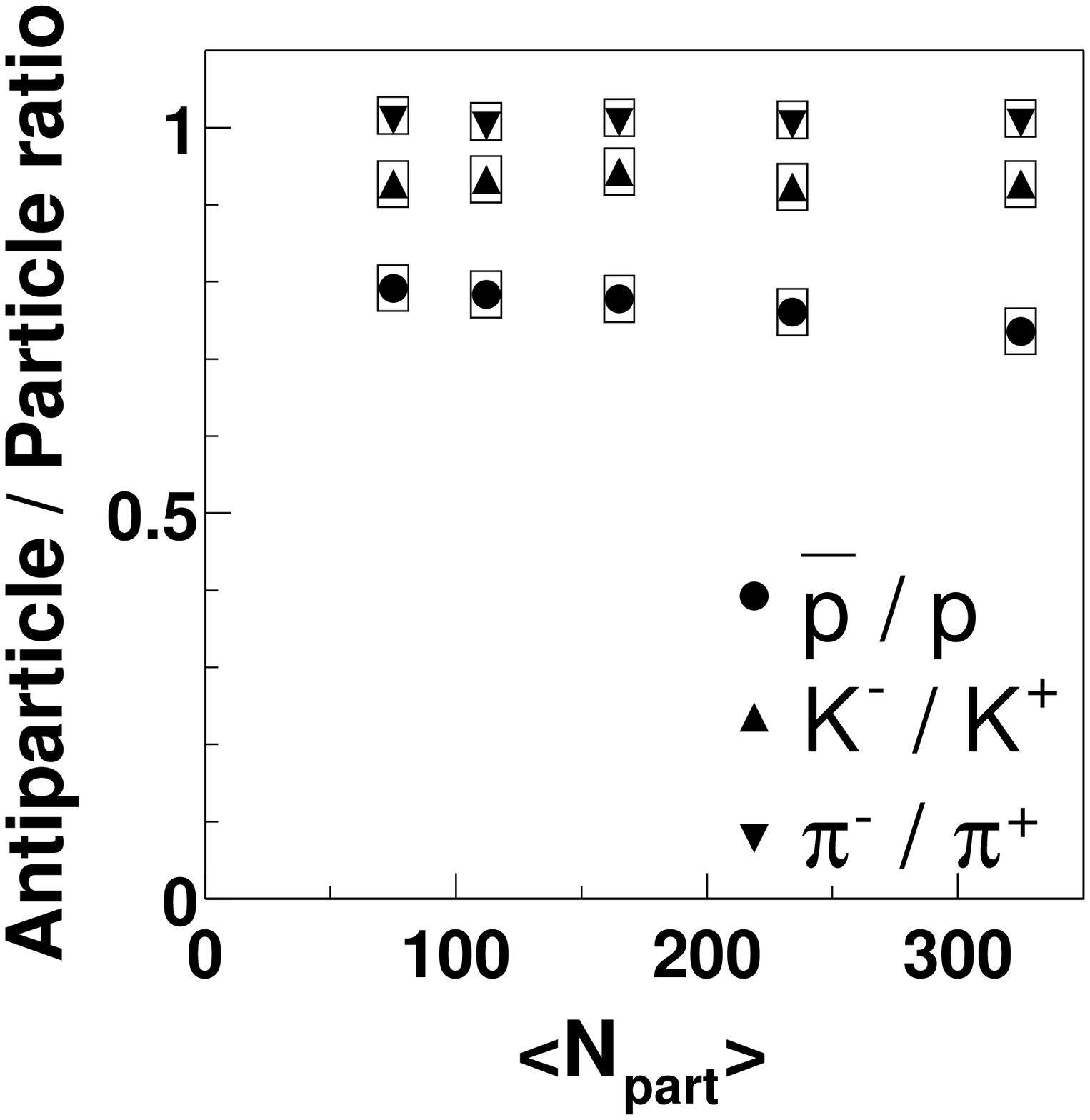,width=65mm}
\caption{Antiparticle to particle ratios as a function of the collision centrality. 
Boxes represent systematic uncertainties.}
\label{fig:toosmall}
\end{minipage}
\end{figure}

Figure 1 shows preliminary results for the dependence of antiparticle
to particle ratios on transverse momentum ($p_T$) for the 10\% most central collisions
at the top RHIC energy. Within our current estimates of the systematic uncertainties, 
the ratios
are independent of $p_T$ in the measured rapidity range. The centrality dependence
of the measured ratios is shown in Fig. 2.  The pion and kaon ratios are consistent
with a constant value over the measured centrality range. For the $<\overline{p}>/<p>$ 
ratio, a
weak but systematic drop with $<N_{part}>$ can be observed. However, within our
uncertainties, the ratio is also consistent with a constant value, independent
of centrality.

\section{YIELDS AT VERY LOW TRANSVERSE MOMENTA}

To extend our identified particle measurements to lower
transverse momenta, we have searched for particles which stop in the fifth 
Si plane of the spectrometer.  The reconstruction
procedure is based on the analysis of tracks with large energy depositions in the
first five spectrometer planes which are located in the field free region of the
spectrometer. 
To determine the particle mass  we check the mass and momentum
hypotheses by making cuts on the energy deposited per unit length in every
plane and the total deposited energy  obtained by summing all the energy depositions. 
The analysis procedure was  tested successfully
on samples of simulated low momentum pions, kaons and protons. 
The measured raw yields were corrected for acceptance and efficiency. The correction
factors were obtained by embedding simulated low momentum particles into
real data events.  Additional corrections,
including feed-down from weak decays and contributions from secondary, misidentified 
and ghost
particles, were also applied.  Our current estimates of the systematic
uncertainties are $\pm$20\% for pions, $\pm$40\% for kaons and $\pm$50\% for protons and 
include
systematic effects in the measured yields and in the
estimated correction factors.

\begin{figure}[htb]
\begin{minipage}[t]{54mm}
%\framebox[79mm]{\rule[-26mm]{0mm}{52mm}}
\epsfig{file=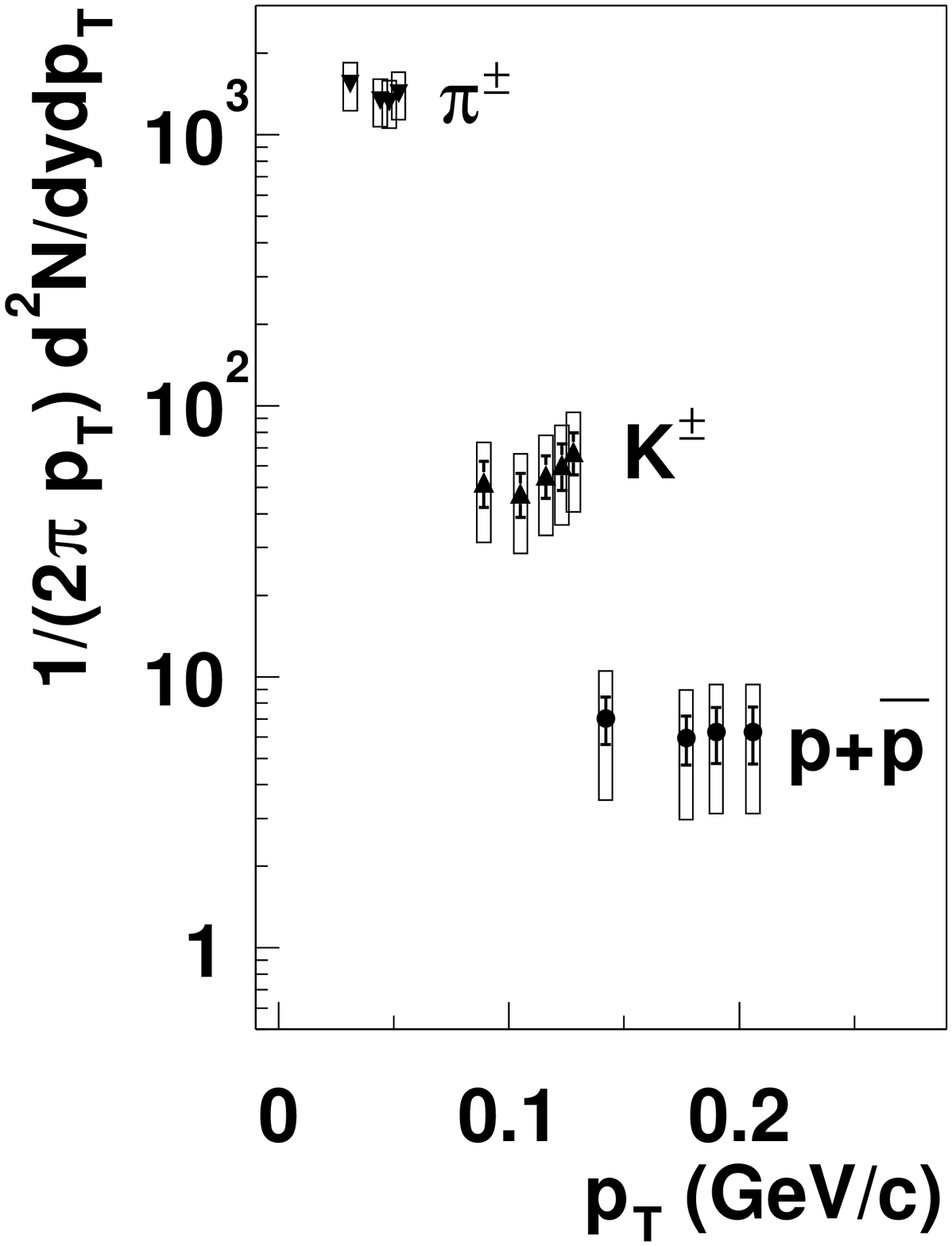,height=72mm}
\caption{Invariant yields as a function of $p_T$. 
The boxes show systematic uncertainties.}
\label{fig:large}
\end{minipage}
\hspace{\fill}
\begin{minipage}[t]{99mm}
%\framebox[74mm]{\rule[-26mm]{0mm}{52mm}}
%\epsfig{file=fig4.eps,width=97mm}
\epsfig{file=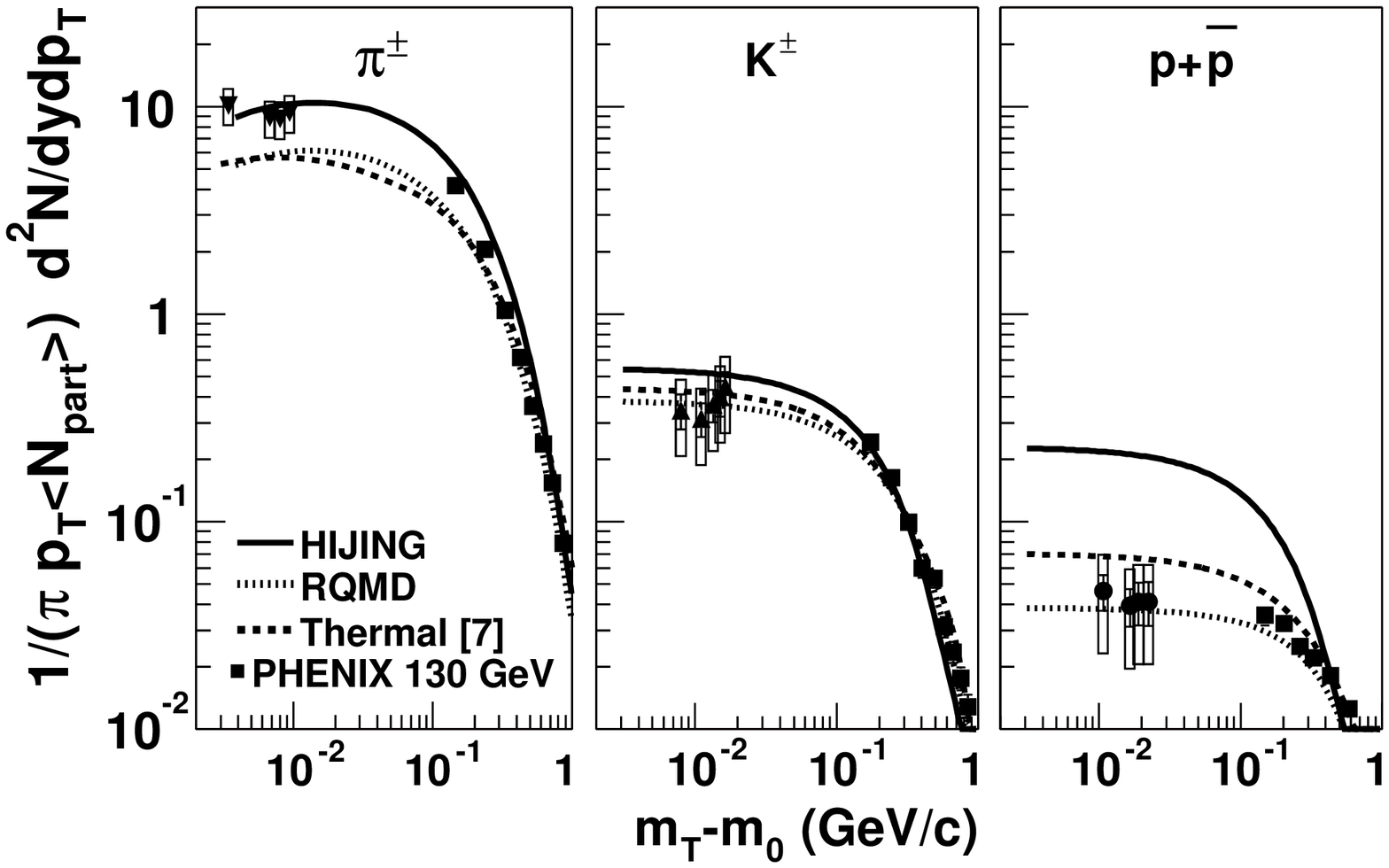,height=72mm}
\caption{Invariant yields divided by $0.5<N_{part}>$ as a function of $m_T - m_0$. Note
the logarithmic horizontal scale.}
\label{fig:small}
\end{minipage}
\end{figure}

Figure 3 shows preliminary results for invariant yields of ($\pi^{\pm}$),
($K^{\pm}$) and ($p + \overline{p}$) measured at mid-rapidity in
the transverse momentum ranges from 30 to 50 MeV/c for charged pions, 90 to
130 MeV/c for kaons and 140 to 210 MeV/c for protons and antiprotons for the 15\% 
most central
Au+Au collisions at $\sqrt{s_{_{NN}}}$ = 200 GeV .
Our results are compared in Fig.4 to the predictions of 
HIJING\cite{hijing}, RQMD \cite{rqmd}
and single freeze-out model of Broniowski and Florkowski \cite{bronflor}. Also shown
in Fig. 4 are PHENIX measurements at 130 GeV \cite{phenix}. All data and the model 
predictions shown in Fig.4 have been corrected for feed-down from weak decays. It is
evident that our measurements explored the range of very low transverse momenta, 
below the current measurements from other RHIC experiments (see also \cite{ullrich}). 
 The yields of ($p + \overline{p}$) seem to agree
with the extrapolation of the PHENIX measurements at higher $p_T$, indicating that
the flattening of proton spectra attributed to the collective radial expansion 
extends down to very low $p_T$. Comparison with the model predictions also
shows that none of the models presented in Fig.4 can consistently describe the pion,
kaon and proton data. Particularly striking differences between the model predictions can be
seen for the ($p+\overline{p}$) yields where the models differ by a factor of 2 to 6.

\section{SUMMARY}

The PHOBOS Collaboration has measured the yields of identified
particles in Au+Au collisions at $\sqrt{s_{_{NN}}}$ = 200 GeV. The $<\overline{p}>/<p>$
ratio reaches a value of 0.73 indicating that only
one quarter of the protons at mid-rapidity have been transported 
from the beam rapidity regions. The estimated baryochemical potential is a factor of 2
smaller at 200 GeV than at 130 GeV. The antiparticle to particle ratios are
consistent with the constant values as a function of transverse momentum and
collision centrality.

 At very low transverse momenta, we see no enhancement in the particle yields.
On the contrary, particle production, particularly
for protons and antiprotons, seems to be suppressed, which can result from a rapid 
transverse expansion
of the system. These measurements provide constraints on models and
suggest that dynamical processes (rescattering, expansion) play an important
role in the description of baryon spectra
at low transverse momenta.

\vspace{1mm}
\noindent
{\small Acknowledgments: 
This work was partially supported by U.S. DOE grants DE-AC02-98CH10886,
DE-FG02-93ER40802, DE-FC02-94ER40818, DE-FG02-94ER40865, DE-FG02-99ER41099, and
W-31-109-ENG-38 as well as NSF grants 9603486, 9722606 and 0072204.  The Polish
group was partially supported by KBN grant 2-P03B-10323.  The NCU group was
partially supported by NSC of Taiwan under contract NSC 89-2112-M-008-024.}

\end{document}